\documentclass[aps,prb,preprint,groupedaddress]{revtex4-2}

\usepackage{graphicx}
\usepackage{times}
\usepackage{amssymb}

\begin{document}

\title{Toward a systematic discovery of artificial functional magnetic materials}

\author{Lukas Botsch}
\affiliation{Division of Superconductivity and Magnetism, Felix-Bloch-Institute
  for Solid State Physics, Universität Leipzig, Linn\'estr. 5, D-04103 Leipzig, Germany}
\email{lukas.botsch@uni-leipzig.de}

\author{Carsten Bundesmann} \affiliation{Tool Development Group, Leibniz
  Institute of Surface Engineering (IOM), D-04318 Leipzig, Germany}

\author{Daniel Spemann} \affiliation{Tool Development Group, Leibniz Institute
  of Surface Engineering (IOM), D-04318 Leipzig, Germany}

\author{Pablo D. Esquinazi} \affiliation{Division of Superconductivity and
  Magnetism, Felix-Bloch-Institute for Solid State Physics, Universität Leipzig,
  Linn\'estr. 5, D-04103 Leipzig, Germany}

\date{\today}

\begin{abstract}
  Although ferromagnets are found in all kinds of technological applications,
  only few substances are known to be intrinsically ferromagnetic at room
  temperature. In the past twenty years, a plethora of new artificial
  ferromagnetic materials have been found by introducing defects into
  non-magnetic host materials. In contrast to the intrinsic ferromagnetic
  materials, they offer an outstanding degree of material engineering freedom,
  provided one finds a type of defect to functionalize every possible host
  material to add magnetism to its intrinsic properties. Still, one
  controversial question remains: Are these materials really technologically
  relevant ferromagnets? To answer this question, in this work the emergence of
  a ferromagnetic phase upon ion irradiation is systematically investigated both
  theoretically and experimentally. Quantitative predictions are validated
  against experimental data from the literature of SiC hosts irradiated with
  high energy Ne ions and own experiments on low energy Ar ion irradiation of
  TiO$_2$ hosts. In the high energy regime, a bulk magnetic phase emerges, which
  is limited by host lattice amorphization, whereas at low ion energies an
  ultrathin magnetic layer forms at the surface and evolves into full magnetic
  percolation. Lowering the ion energy, the magnetic layer thickness reduces
  down to a bilayer, where a perpendicular magnetic anisotropy appears due to
  magnetic surface states.
\end{abstract}

\maketitle

\section{Introduction}
Magnetic materials play a major role in many spintronic and other technological
applications~\cite{Hirohata2020} such as magnetic storage~\cite{Chappert2007},
logic devices~\cite{Sharma2020}, magnetic field sensors and magnetic random
access memory~\cite{Bhatti2017,Song2017,Meena2014}. Materials with strong
intrinsic ferromagnetic (FM) order above room temperature, such as the
transition metals Fe, Ni or Co and their alloys, are rather unusual among the
magnetic materials known today~\cite{Connolly1972} and there is still the need
for new functional materials with magnetic order above room temperature. In the
past two decades, a method of creating artificial ferromagnetic materials has
emerged and a multitude of so-called defect-induced ferromagnets were
reported~\cite{Ning2015,Esquinazi2020,Wang2020}. Since the first prediction of
an artificial ferromagnetic material with transition temperature above 300~K,
based on Mn doped ZnO appeared twenty years ago~\cite{Dietl2000}, the field has
substantially evolved. First, it was realized that doping with magnetic
impurities was not at all necessary in order to induce a robust FM order in the
non-magnetic host matrix, rather all kinds of lattice defects were at the origin
of the measured magnetic
signals~\cite{Sundaresan2006,Khalid2009,Hong2007,Hong2008}. This realization
promised great possibilities to construct new functional magnetic materials, as
any non-magnetic material could potentially host a certain kind of defect,
turning it into an artificial ferromagnet. The hunt was on and the result was a
plethora of reports ranging from oxide, nitride, carbon-based, 2D van der Waals
and many more materials showing signals of ferromagnetism upon introducing all
kinds of defects~\cite{Ning2015,Esquinazi2020,Wang2020}.
One of the most promising and versatile methods for introducing these defects is
the irradiation with non-magnetic ions~\cite{Zhou2014}, owing to the
availability of ion sources ranging over the whole periodic table and energies
from few eV to hundreds of MeV.

Although many experiments were accompanied by theoretical studies, such as
electronic structure calculations based on density functional theory (DFT), the
search was mostly guided by blind trial and error and a brute force approach. It
is therefore not very surprising that most of the reported materials only showed
very tiny magnetic signals, which soon led to debates about the nature of the
effect~\cite{Ackland2018,Coey2019} and raised the question of whether this route
could eventually lead to a robust magnetic order above room temperature,
comparable with intrinsic ferromagnets. Furthermore, the measurement of the
magnetization of such artificial ferromagnetic samples turns out to be quite
difficult due to the inherent uncertainty of the magnetic volume, leading to
largely underestimated values in the literature. Considering the enormous amount
of host material candidates and lattice defects, a more systematic search method
and better selection criteria are highly needed.

In this work, we present a systematic investigation of the emergence of such
artificial ferromagnetic phases, both theoretically and experimentally. We first
propose a scheme for the computational discovery of candidate materials that can
be created by ion irradiation. The scheme is based on first principle
calculations, guided by experimental constraints, automatically restricting the
potential defects to those accessible experimentally and can readily be
implemented for high throughput material discovery. We further provide a method
to determine the defect distribution created within the host lattices, allowing
to obtain accurate magnetization values.

Two ion energy regimes are then investigated, namely at high energies $\geq
100$~keV and at low energies $\leq 1$~keV. The main physical processes governing
the emerging FM phase in these regimes are identified and validated using high
energy experimental data found in the literature and own low energy experiments.
The predictions of the scheme are compared to experimental magnetization data of
SiC samples irradiated with high energy Ne ions, found in the literature. The
comparison confirms the role of host lattice amorphization as a limiting factor
in the magnetic percolation at high ion energy.

We then present own systematic experimental results, showing the emergence of an
artificial FM phase in TiO$_2$ hosts, upon irradiation with low energy ($\leq
1$~keV) Ar ions. As predicted by our simulations, the magnetic percolation is
most affected by sputtering processes at the surface in the low ion energy
regime. The FM phase emerges in an ultrathin region beneath the surface, whose
thickness varies from a few atomic layers down to a magnetic bilayer, depending
on the ion energy. In these ultrathin films, the emerging FM phase reaches the
full magnetic percolation limit, where it spans over the complete sample surface
area. The magnetic anisotropy is also investigated, showing a switch from
in-plane to out-of-plane easy magnetization direction as the ion energy and the
resulting magnetic layer thickness decreases. This phenomenon is explained by
the contribution of the surface to the magnetocrystalline anisotropy.

\section{Computational Methods}

Most of the theoretical work related to artificial ferromagnets has so far been
devoted to understanding the origin of the magnetic signals observed
experimentally in different host systems. Guided by experimental intuition, a
considerable computational effort was undertaken to identify possible defect
structures, that carry non-zero magnetic moment and could explain the FM signals
measured in nominally non-magnetic host systems upon ion irradiation. The method
of choice are spin-polarized electronic structure calculations performed on
different levels of DFT, which yield the magnetic ground state of the defective
systems and can provide an estimate of the magnetic percolation threshold. This
is the threshold density of defects needed in a certain host matrix for a long
ranged ordered FM phase to emerge. Most studies rely on supercell methods to
model systems with different defect concentrations. As the number of possible
defect structures that could potentially yield a magnetic ground state in each
single host matrix is enormous, studies are usually limited to investigating
simple point defects, such as vacancies or interstitials. Obviously, an
exhaustive search through all possible defect structures is not practical and a
better method is needed.

Considering artificial ferromagnets created by ion irradiation, a much better
starting point would be to only consider those defect structures that are
experimentally accessible, i.e. that are likely to be created by the impact of
energetic ions on the host matrix. Calculations of irradiation damage have been
a standard tool in the context of accelerator physics for a long time. Molecular
dynamics (MD) methods, taking into account different levels of interatomic
potentials, exist and yield accurate simulations of the damage resulting from
collision cascades in a wide range of energies~\cite{Nordlund2018}. The
resulting damage structures, calculated in large host systems of several
thousand atoms, can then be decomposed into smaller units of equivalent defects
using cluster analysis (CA) techniques. These simulations not only yield
structures of potential defect complexes arising in the collision cascades,
beyond the simple vacancy or interstitial, but can also give statistical
information about their creation probabilities.

\begin{figure}
  \centering
  \includegraphics[width=5.5cm]{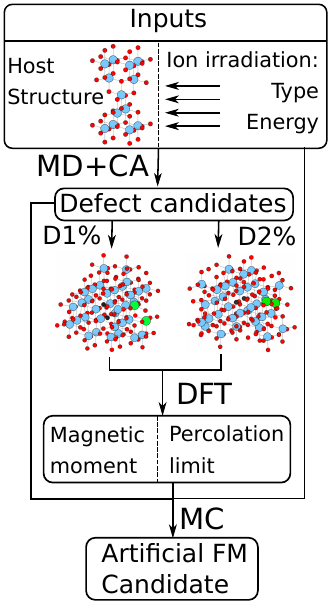}
  \caption{\label{fig:1} Computational scheme for the prediction of artificial
    functional magnetic materials. It takes as inputs the atomic structure of a
    host material and the ion type and energy range of the ion irradiation. In a
    first step, molecular dynamics (MD) simulations and cluster analysis (CA)
    algorithms yield possible defect structures and their corresponding creation
    probabilities, likely to be formed during the ion irradiation process. The
    resulting defective structures are used as input for DFT electronic
    structure calculations, giving the magnetic ground state and the percolation
    limit for each defect type. Finally, using the results of all calculations
    and applying Monte Carlo (MC) methods, a magnetic phase diagram is
    constructed, which indicates the irradiation parameters likely to create an
    artificial FM. }
\end{figure}

Taking the resulting structures of the defective host material as input for
spin-polarized DFT calculations, allows us to save substantial computational
effort and gives much more realistic results. Building on these remarks, we
propose the computational scheme depicted in Figure~\ref{fig:1}, taking as input
the atomic structure of a host material, the type and energy of the ion
irradiation, from which potential defective structures and their creation
probabilities are obtained using MD simulations and CA algorithms. DFT
electronic structure calculations are then performed for the resulting defective
structures and their magnetic ground state is determined. For defects yielding a
non-zero magnetic moment, the percolation limit is estimated. Finally, taking
into account ion energy loss in the irradiated host material and the defect
formation probabilities predicted by the MD simulations, a magnetic phase
diagram can be constructed using Monte Carlo (MC) methods. From this phase
diagram, quantitative predictions of the total magnetic moment, the magnetic
volume and magnetization can be extracted.

\section{High Energy Ion Irradiation in S\MakeLowercase{i}C}

To validate the predictions of the computational scheme, we first calculate the
magnetic phase diagram of 6H-SiC, resulting from the irradiation with high
energy ($E_\mathrm{ion} = 140$~keV) Ne ions and compare the predicted
magnetization values with those measured experimentally and published by Li and
coworkers~\cite{Li2011}. In the following sections we aim to give a step-by-step
example of the calculations involved in constructing the magnetic phase diagram
and extracting quantitative predictions for the emerging FM phase.

\subsection{Molecular Dynamics Collision Cascade Simulations}

\begin{figure}
  \centering
  \includegraphics[width=5.5cm]{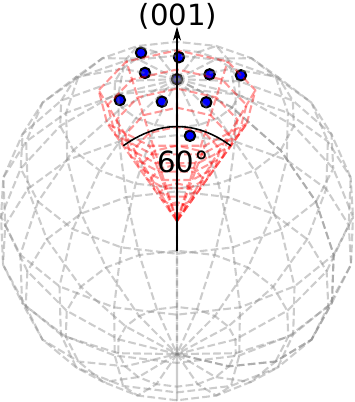}
  \caption{\label{fig:2} PKA directions sampled in the conical cut of the unit
    sphere, with aperture $60^\circ$ and axis along (001) crystal direction.
    Blue dots indicate the 10 initial directions used in the collision cascade
    simulations. }
\end{figure}

To find the defects produced in 6H-SiC resulting from high energy ion
irradiation, we performed a total of 9600 collision cascade simulations in
6H-SiC using the LAMMPS MD code~\cite{lammps,Plimpton1995,Brown2012}. The
interatomic interactions were modeled using Tersoff/ZBL empirical potentials, as
described by Devanathan et al.~\cite{Devanathan1998} and used previously for
similar simulations~\cite{Li2019}. To compute the collision cascades, systems of
$20 \times 20 \times 20$ unit cells (corresponding to 96000 atoms) with periodic
boundary conditions were constructed. The unit cell parameters were set as
$a=b=3.095$~\AA, $c=15.185$~\AA, $\alpha=\beta=90^\circ$, $\gamma=120^\circ$.
Twelve initial structures were equilibrated in the microcanonical (NVE) ensemble
for 10-21~ps, respectively, with a timestep of 1~fs at $T=300$~K. One Si atom
and one C atom located at the center of the simulation cells were selected as
primary knock-on atoms (PKA). Their initial kinetic energy was then set to
values in the range 5~eV-200~eV by fixing the initial velocity along 10
different directions sampled from a cone with main axis along the (001) crystal
direction and an aperture of $60^\circ$ (see Figure~\ref{fig:2}).

After setting the initial kinetic energy of the PKA atom, the systems were let
to evolve in three phases, with timesteps of 0.01~fs, 0.1~fs and 1~fs for
0.1~ps, 1~ps and 10~ps, respectively, in order to capture the whole ballistic
dynamics of the collision cascades.

\begin{figure}
  \centering
  \includegraphics[width=10.5cm]{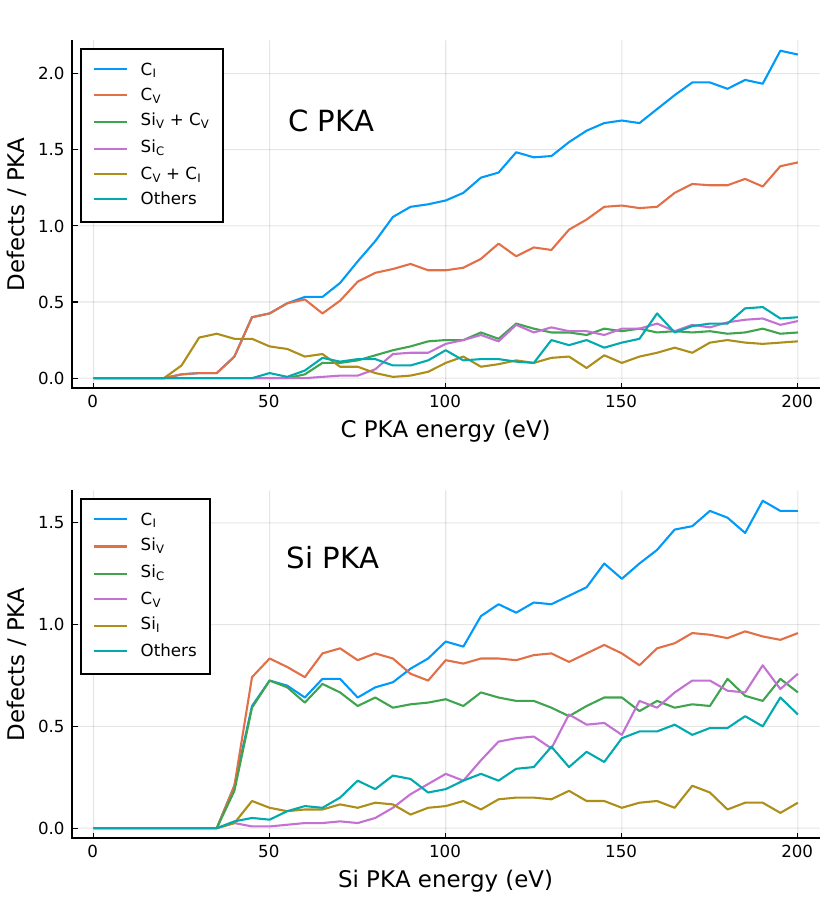}
  \caption{\label{fig:3} Defect creation rates in 6H-SiC obtained from collision
    cascades simulated by molecular dynamics methods for C and Si PKAs in the
    energy range 5~eV to 200~eV. The five most prevalent defect types are shown,
    all other defects are grouped together. C$_I$: carbon interstitial; C$_V$:
    carbon vacancy; Si$_I$: silicon interstitial; Si$_V$: silicon vacancy;
    Si$_\mathrm{C}$: silicon antisite.
  }
\end{figure}

The resulting collision cascade trajectories were then analyzed using the Ovito
library. Single point defects (vacancies, interstitials, antisites) were
identified using a Wigner-Seitz decomposition of the initial and final
structure. Defect clusters were then identified using a clustering algorithm
with a length cutoff of 2.5~\AA, yielding the number of defect types created
during each of the simulated collision cascades. The large degree of statistical
sampling (120 simulations per PKA type and energy) allowed us to determine
average defect creation rates, which are shown in Figure~\ref{fig:3} for the C
and Si PKAs in the energy range 5~eV to 200~eV. We find a displacement threshold
of $E_d = 25$~eV and $E_d = 40$~eV for the C and Si PKA, respectively, in
agreement with previous reports~\cite{Li2019}.

The most prevalent defects are the C interstitial (C$_\mathrm{I}$), C vacancy
(C$_\mathrm{V}$), Si vacancy (Si$_\mathrm{V}$), Si antisite (Si$_\mathrm{C}$)
and the di-vacancy (Si$_\mathrm{V}$ + C$_\mathrm{V}$). We note that the creation
rates shown in Figure~\ref{fig:3} are given as average number of defects created
per PKA event of a certain kinetic energy and type. In general, incident ions
collide with one or more lattice ions, transferring some of their kinetic energy
to the PKA. For a complete picture of the defect creation process, we need to
determine the energy and spacial distribution of PKA events, created by
energetic ions. But first, we need to identify those defects that carry non-zero
magnetic moment and determine the sign and range of their exchange interactions.

\subsection{Spin properties of the di-vacancy and Si vacancy in 6H-SiC}

The spin properties of both the di-vacancy and the Si-vacancy in SiC have been
extensively investigated due to their potential use as spin-defect qubits, both
theoretically~\cite{Torpo1999, Liu2011, Wang2015b, Wang2015c, Soykal2016,
  Davidsson2019} and experimentally~\cite{Soerman2000, Janzen2009, Li2011,
  Wiktor2014, Wang2015b, Wang2015c, Christle2017, Davidsson2019, Pavunny2021}.
According to spin-density functional theory calculations, the spin states and
interactions strongly depend on the charge state of the defects~\cite{Torpo1999,
  Wiktor2014, Liu2011, Wang2015c}. While the negatively charged Si vacancy has a
spin-3/2 ground state, the neutral and higher charge states are not
spin-polarized~\cite{Torpo1999,Janzen2009}. The neutral and negatively charged
di-vacancies have a spin-1 ground state~\cite{Son2006, Liu2011, Wang2015c}. In
both defects, the calculated spin densities indicate that the polarization
originates from localized states at the C atoms surrounding the Si vacancy.

Wang et al. investigated the exchange coupling between charged di-vacancies as a
function of the defect distance~\cite{Wang2015c} and found that the negatively
charged di-vacancies couple ferromagnetically at distances 10-18.5~\AA,
corresponding to a percolation threshold of $\sim1$ at.\%.

\subsection{Binary Collision Monte Carlo Simulations}

Out of all defects predicted in the collision cascade simulations, the Si and
di-vacancy are the most promising candidates to induce a long-ranged
magnetically ordered phase in 6H-SiC, as they both carry non-zero magnetic
moment, tend to couple ferromagnetically and both Si and C vacancies are created
with a high probability (see Figure~\ref{fig:3}) in the collision cascades.

In order to relate the defect creation rates obtained from the MD collision
cascade simulations to ion irradiation experiments, we need to determine the
energy and spacial distribution of PKA events resulting from the collision of
high energy ions with the SiC lattice atoms. We obtain this using
SRIM~\cite{Ziegler2008,Ziegler2010}, a widely used binary collision Monte Carlo
code, that simulates the stopping and range of ions in matter by assuming the
target material is amorphous. In the full collision simulation mode, SRIM takes
as input the number of ions to simulate, their type and kinetic energy, the
target material composition and density, the displacement threshold of each
target atom type and outputs a list of PKA collision events, including the
transferred kinetic energy and the position.

\begin{figure}
  \centering
  \includegraphics[width=10.5cm]{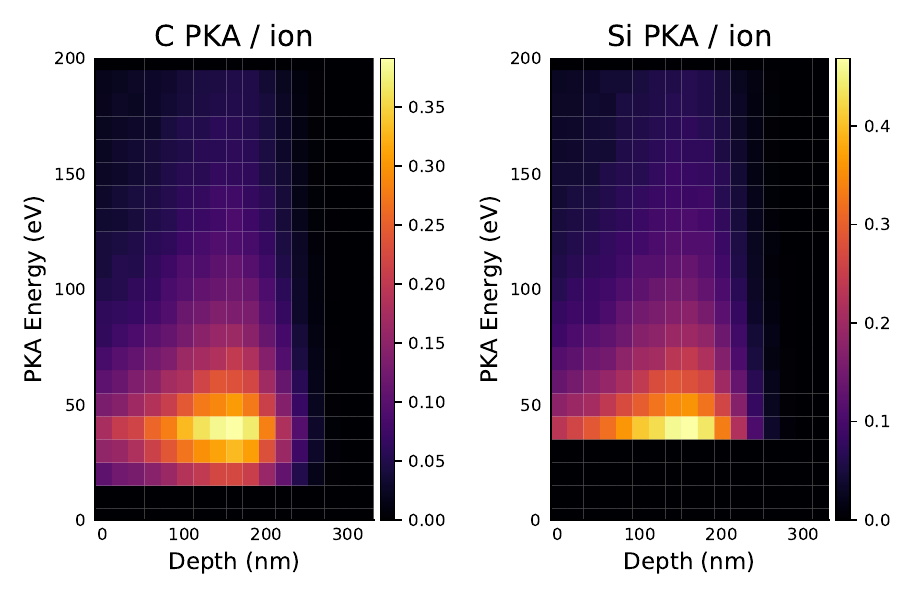}
  \caption{\label{fig:4} PKA event distribution for 6H-SiC irradiated with Ne
    ions at $E_\mathrm{ion} = 140$~keV, calculated by binary collision Monte
    Carlo simulations using SRIM. The number of resulting primary knockon atom
    (PKA) events per incident ion is shown on the color scale (right axis) for C
    and Si PKAs as a function of the depth along the (001) crystal direction
    (horizontal axis) and the PKA energy (left axis). }
\end{figure}

To be able to compare our computational predictions to the experimental data
published by Li and coworkers~\cite{Li2011}, we performed SRIM simulations of
$10^{5}$ Ne$^+$ ions at $E_\mathrm{ion} = 140$~keV, incident on amorphous SiC,
setting the displacement thresholds of the C atoms to $E_d = 25$~eV and that of
Si to $E_d = 40$~eV, according to our MD simulations (see Figure~\ref{fig:3}).
We analyzed the resulting PKA events using a histogram method with a bin size of
5~eV for the PKA energy, matching the energy resolution of our MD simulations.
The PKA energy and depth (along the irradiation axis) distribution is shown for
the two PKA types in Figure~\ref{fig:4}. In both cases, the maximum of the
distribution lies around 40-50~eV and at a depth of 160~nm below the surface.
The maximum range of the Ne$^+$ ions is 260~nm. We note that, even though the
initial kinetic energy of the ions was set to 140~keV, more than 90~\% of the
PKA events occur at an energy below 200~eV, indicating that the ions transfer
most of their kinetic energy to the electronic system of the target material
before undergoing binary collisions with the target nuclei.

\subsection{Quantitative predictions and experimental validation}

\begin{figure}
  \centering
  \includegraphics[width=8.5cm]{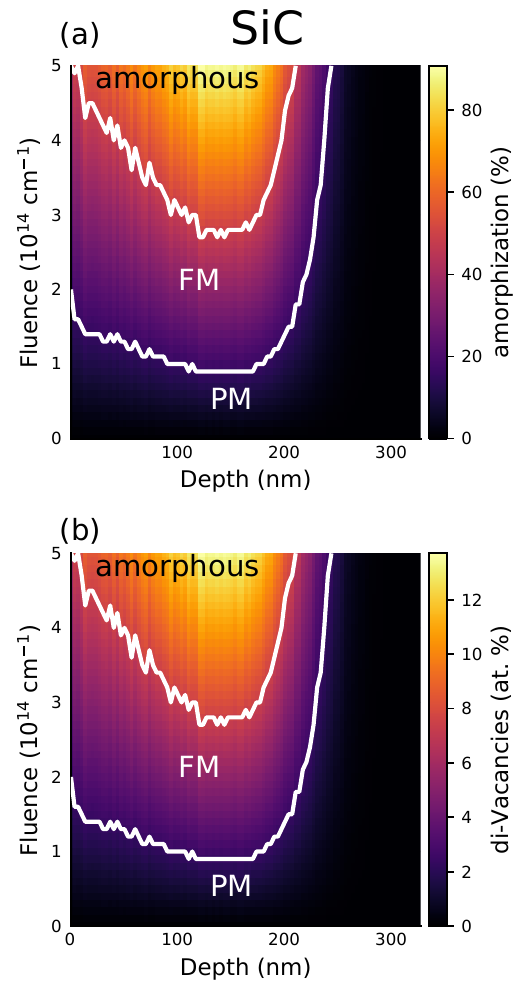}
  \caption{\label{fig:5} Magnetic phase diagram of 6H-SiC, irradiated with Ne
    ions at $E_\mathrm{ion} = 140$~keV. FM: ferromagnetic, PM: paramagnetic. The
    degree of amorphization (a) and concentration of di-vacancies (b) are shown
    on the color scales (right axis) and as a function of depth along the (001)
    crystal direction (horizontal axis) and the irradiation fluence (left axis).
  }
\end{figure}

Combining the results of the MD and SRIM simulations, we calculated the
densities of all defect types found in the MD simulations, as a function of ion
fluence in the range 0-$5\times 10^{14}$~cm$^{-2}$ (see Figure~\ref{fig:5}).
Pairs of single C and Si vacancies, created in close proximity were counted
towards the di-vacancy density (Figure~\ref{fig:5}(b)) and the degree of
amorphization (Figure~\ref{fig:5}(a)) was defined as the total number of defects
per lattice atom. As the creation rate of C vacancies is significantly larger
than that of Si vacancies, the resulting density of isolated Si vacancies is
negligible and we will therefore only take into account the di-vacancy defect in
our further discussion.

Taking into account a percolation threshold of 1 at.\% for the di-vacancy
defect~\cite{Wang2015c}, we identify the threshold ion fluence required to
induce long-ranged FM order along the irradiation direction, as indicated in
Figure~\ref{fig:5} by the lower white line. Below this line, at low ion
fluences, the defect density is too low to create an artificial FM phase and the
material is paramagnetic. Above the line, at high enough fluences, a magnetic
percolation transition occurs and a FM phase emerges.

As the spin polarization of the di-vacancy defect originates from the C atoms
surrounding the Si vacancy, it appears plausible that a high degree of
amorphization could destroy the FM phase. We have therefore calculated a second
threshold fluence, at which the degree of amorphization along the irradiation
direction reaches 50~\%, as indicated by the upper white line in
Figure~\ref{fig:5}. Above this threshold, at least two of the four C atoms
surrounding the Si vacancy are displaced on average.

\begin{figure}
  \centering
  \includegraphics[width=8.5cm]{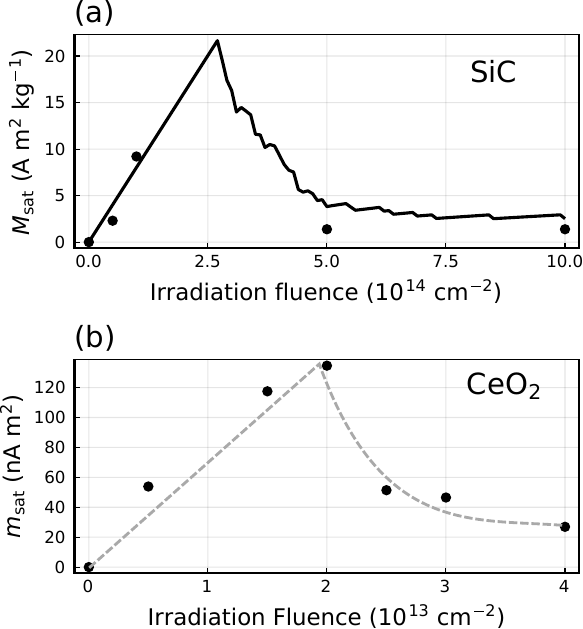}
  \caption{\label{fig:6} (a) Saturation magnetization of the ferromagnetic
    signal as a function of the ion irradiation fluence in a 6H-SiC single
    crystal sample irradiated with Ne$^+$ ions at $E_\mathrm{ion} = 140$~keV
    (data reproduced from~\cite{Li2011}); (b) Saturation magnetic moment of a
    CeO$_2$ bulk sample, irradiated with Xe$^+$ ions at $E_\mathrm{ion} =
    200$~MeV (data reproduced from~\cite{Shimizu2012}). The magnetic properties
    were measured after each irradiation step. The solid line in (a) show the
    calculated magnetization as described in the text; in (b) the line is just a
    guide for the eye. All measurements were done at room temperature.}
\end{figure}

These two boundaries define a volume, in which a FM phase emerges. By
integrating over all defect magnetic moments in this volume, we obtain the
saturation magnetization of the FM phase. This is shown as a solid line in
Figure~\ref{fig:6}(a), as a function of the irradiation fluence. Up to a fluence
of $2.5\times 10^{14}$~cm$^{-2}$, the magnetization increases linearly, as more
defects are created. At higher fluences, the amorphization threshold is reached
in a large portion of the magnetic volume and the magnetization rapidly
decreases.

In order to compare our theoretical predictions with the experimental data
published by Li et al.~\cite{Li2011}, we calculated the saturation magnetization
from the total magnetic moment data measured using SQUID magnetometry by using
the sample area and the depth of the magnetic phase resulting from our
simulations. The resulting magnetization values are indicated in
Figure~\ref{fig:6}(a) as bullets. The magnetization observed experimentally
first increases with the Ne ion fluence and decreases at large enough fluences,
matching our theoretical predictions quantitatively.

The authors suspected that the amorphization could play a role in the
disappearance of the FM signal in their samples at larger
fluences~\cite{Li2011}. Our simulations confirm that the amorphization of the
host lattice is the major limiting factor for the development of a dense FM
phase in SiC. This might also be the case in other materials, where artificial
FM phases have been created using ion irradiation. Detailed experimental data
showing the evolution of such a FM phase as a function of the ion fluence is
scarce. Shimizu et al. measured the magnetic properties of CeO$_2$ single
crystals irradiated with Xe$^+$ ions at $E_\mathrm{ion} = 200$~MeV using SQUID
magnetometry~\cite{Shimizu2012}. Figure~\ref{fig:6}(b) shows the evolution of
the total saturation magnetic moment of an emerging FM phase in CeO$_2$, as a
function of the ion fluence. There, the magnetic moment follows the same
qualitative trend: It first increases rather linearly and at a fluence $>
2\times 10^{13}$~cm$^{-2}$, it decreases.

\section{Low Energy Ion Irradiation in Anatase T\MakeLowercase{i}O$_2$}

Most experimental investigations of artificial FM phases emerging upon ion
irradiation reported in the literature were performed at high ion energies $\geq
100$~keV. This results in a rather large ion penetration depth and as we showed
in the previous section, the evolution of the artificial FM phase is limited by
the amorphization of the host lattice. In this section, we present results of
our systematic experimental investigation of a FM phase emerging in anatase
TiO$_2$ hosts upon low energy Ar$^+$ ion irradiation ($E_\mathrm{ion} \leq
1$~keV).

\subsection{Experimental Methods}

\begin{figure}
  \centering
  \includegraphics[width=8.5cm]{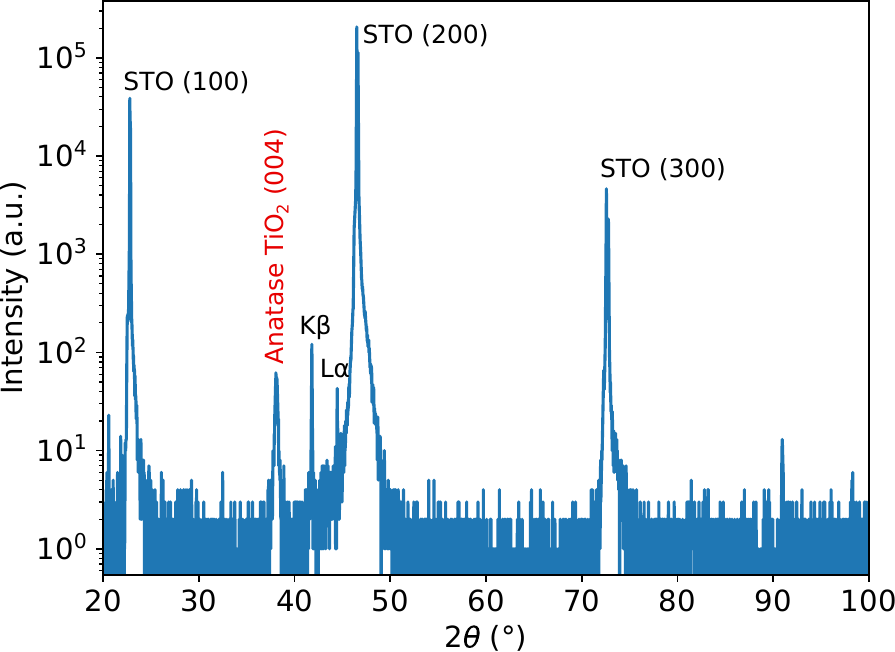}
  \caption{\label{fig:7} X-ray diffraction measurements of the crystallized
    anatase TiO$_2$ films grown on SrTiO$_3$ (100) substrates. The data shown
    here were obtained from sample S1000 after the last irradiation step. }
\end{figure}

Amorphous TiO$_2$ thin films were grown on SrTiO$_2$ substrates by ion beam
sputter deposition~\cite{Bundesmann2018}. Here a beam of low-energy Ar ions is
directed onto a Ti target. Due to momentum and energy transfer, target particles
get sputtered and condense on a substrate. Additionally oxygen background gas
was provided such that TiO$_2$ thin films were formed. Ion energy, ion current,
and ion incidence angle were 1000~eV, 7~mA, and 30$^\circ$, respectively. The
substrates were placed at a polar emission angle of 40$^\circ$ relative to the
target normal. The volumetric flow rate of Ar primary gas was 3.5~sccm and of
O$_2$ background gas 2.0~sccm, which resulted in a total working pressure of $6
\times 10^{-3}$~Pa. More details are given in Ref.~\cite{Bundesmann2017}. The
films have a thickness of 40~nm. After annealing at 500$^\circ$~C in air for 1h,
the films crystallize in the anatase phase with the film surface normal along
the (001) crystal direction, as confirmed by XRD measurements (see
Figure~\ref{fig:7}).

Three thin film samples with a surface area of $5 \times 5$~mm$^2$ were selected
for the ion irradiation experiments. Each sample was then irradiated with Ar$^+$
ions at $E_\mathrm{ion} = 200$~eV (``S200''), 500~eV (``S500'') and 1000~eV
(``S1000''), respectively, using a custom-built DC plasma chamber. The ion
current was measured through a gold frame surrounding the sample during the
irradiation process and was used to calculate the fluence.

The magnetic properties of each sample were characterized before and after each
irradiation step using a commercial SQUID magnetometer (Quantum Design MPMS XL)
using the reciprocating sample operation (RSO) mode. In order to ensure the
comparability of the results after each irradiation step, great care has been
taken to reduce as much as possible any source of contamination to the samples
and the same protocol and schedule was maintained throughout the experiment. The
samples were clamped in a plastic straw (as shown in Figure~10(b)
in~\cite{Sawicki2011}), allowing to measure the magnetic responses to a magnetic
field applied perpendicular and parallel to the film surface. The total magnetic
moment of the sample was recovered from the raw SQUID voltage signal using a
point dipole approximation. The finite sample size has been corrected for using
methods described in Ref.~\cite{Sawicki2011}, by applying a correction factor to
the total moment (in-plane: 0.969296, perpendicular: 1.015966). For magnetic
hysteresis loop measurements, $m(B)$, the magnetic field was first reduced from
0.1~T to nominally zero in oscillating mode, followed by a magnet quench to
minimize the remanent field. In addition, the solenoid hysteresis was corrected
for using a calibration measurement, as described in detail
in~\cite{Sawicki2011}. For temperature dependent measurements, $m(T)$, the
sample temperature was first set to $T=380$~K, followed by the same magnetic
field reset procedure. After cooling the sample down to $T=2$~K, a constant
magnetic field was set and the zero field cooled (ZFC) curve was measured while
heating up to $T=380$~K followed by the field cooled (FC) measurement. At each
temperature step, we waited for 60~seconds to allow the sample to reach thermal
equilibrium prior to the measurement.

To recover the relevant magnetic hysteresis parameters from the experimental
data, we use the following model:
\begin{equation}
  \label{eq:hysteresis_model}
  m(B) = \chi B + \frac{2m_s}{\pi} \arctan\left[ \frac{B \pm B_c}{B_c} \tan\left( \frac{\pi m_r}{2m_s} \right) \right],
\end{equation}
where $\chi$ accounts for a linear contribution to the susceptibility, including
a diamagnetic response from the substrate and a paramagnetic response at room
temperature and moderate fields, $m_s$, $B_c$ and $m_r$ are the saturation
moment, coercive field and remanent moment of a hysteretic response.

After the measurements of sample S500, and before measuring the other two
samples, we had to adjust the ``SQUID tuning parameters'' (drive power and
frequency) as the SQUID was slightly detuned, resulting in a significant
reduction of the signal noise.

\subsection{Magnetic phase diagram}

To understand the emergence of an artificial FM phase in TiO$_2$ due to low
energy ion irradiation, we used the same computational scheme as for the high
energy case, outlined in the previous section. Robinson et
al.~\cite{Robinson2014} performed detailed molecular dynamics simulations of
collision cascades in anatase TiO$_2$, due to low energy ion irradiation and
calculated the probability of resulting defect structures, as shown in
Figure~\ref{fig:2}, for collision cascades resulting from Ti and O primary
knock-on atoms (PKA), using a very similar method to ours. At PKA energies near
the displacement threshold, $E_d=39$~eV for Ti PKAs and $E_d=19$~eV for O, the
primary defects created are the di-Frenkel pair (dFP; 40\% of Ti PKAs),
consisting of two Ti atoms displaced into interstitial sites leaving behind two
vacancies and the oxygen vacancy (O$_\mathrm{v}$; 50\% of O PKAs) (see
Figure~\ref{fig:2}(a,c)).

\begin{figure}
  \centering
  \includegraphics[width=12.5cm]{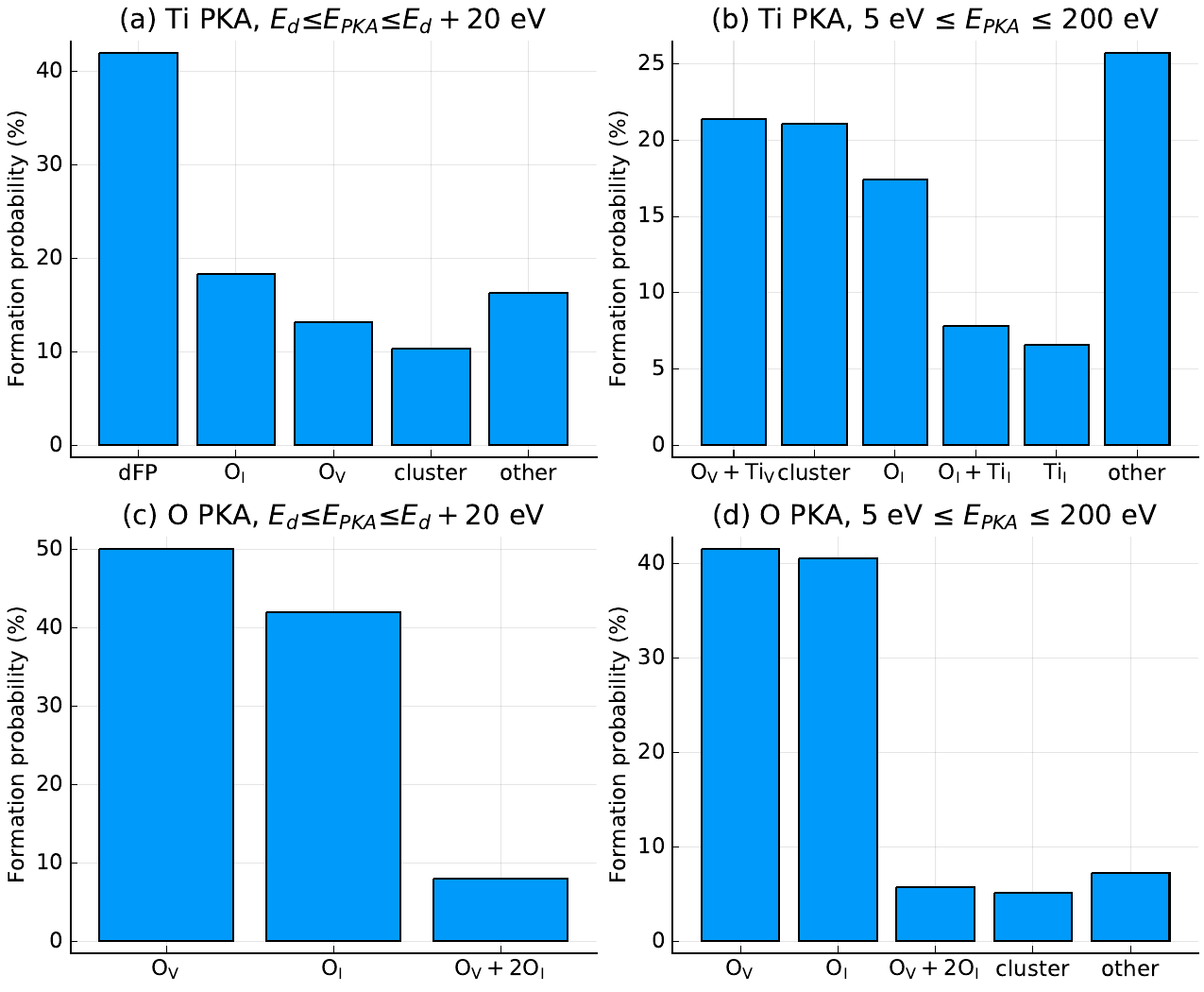}
  \caption{\label{fig:8} Defect formation probabilities at primary knockon (PKA)
    energies $E_\mathrm{PKA}$ near the displacement threshold $E_d$ (a,c) and in
    the range $5-200$~eV (b,d) for Titanium (a, b) and Oxygen (c, d) PKAs.
    Complexes containing more than four defects are categorized as ``cluster'',
    the group labeled ``other'' contains defects with less than 5\% formation
    probability. Data taken from Ref.~\cite{Robinson2014}. }
\end{figure}

Both the dFP~\cite{Stiller2020} and O$_\mathrm{v}$~\cite{Li2007} defects in
anatase TiO$_2$ have been investigated previously using DFT calculations and
found to carry a magnetic moment of $2\mu_B$ and $1\mu_B$, respectively. When
their concentration reaches $\sim 5$\% in the bulk, the magnetic moments start
to couple ferromagnetically and undergo a magnetic percolation transition. A
long-ranged ordered phase emerges and persists even above room
temperature~\cite{Stiller2020}. According to reports from Stiller et al., dFP
defects are mostly responsible for the emergence of a FM phase in anatase
TiO$_2$ irradiated with low energy Ar$^+$ ions~\cite{Stiller2020}. The
O$_\mathrm{v}$ defect was found to play a role in magnetic TiO$_2$ doped with
Cu~\cite{Li2007}.

\begin{figure}[h!]
  \centering
  \includegraphics[width=13.5cm]{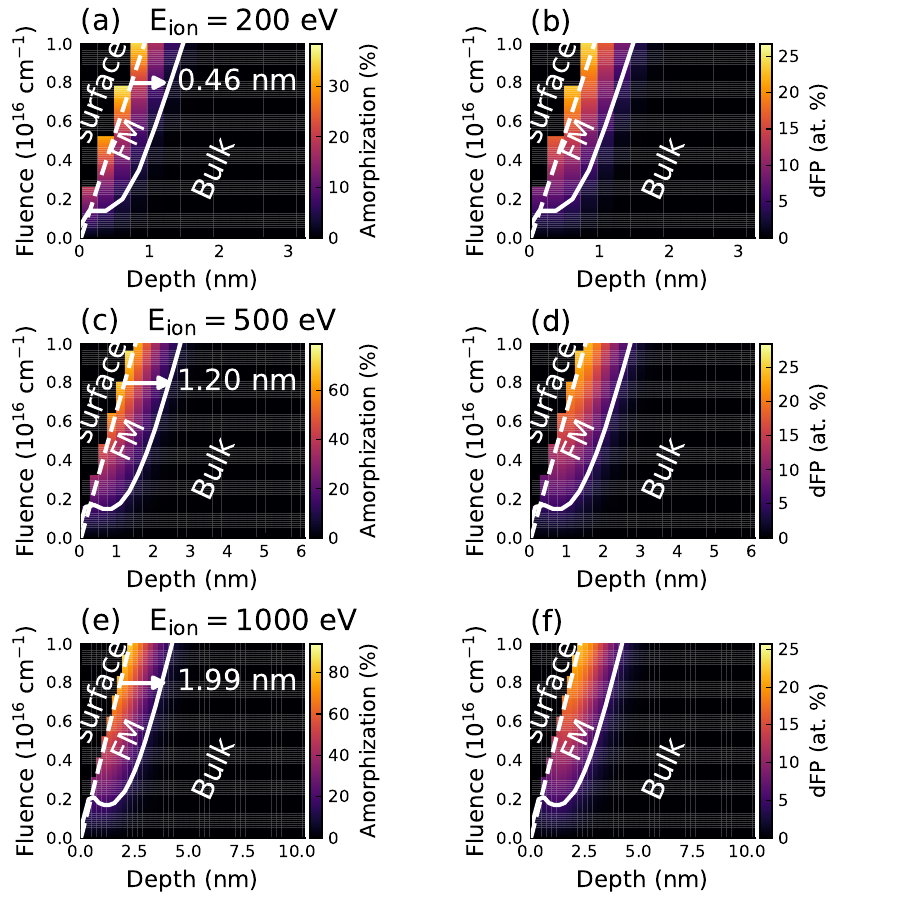}
  \caption{\label{fig:9} Magnetic phase diagram for the artificial FM created in
    anatase TiO$_2$ by Ar$^+$ ions at $E_\mathrm{ion} = 200$~eV (a,b),
    $E_\mathrm{ion} = 500$~eV (c,d) and $E_\mathrm{ion} = 1000$~eV (e,f). The
    color scale (right axes) shows the degree of amorphization (a,c,e) and the
    density of dFP defects (b,d,f), along the irradiation direction (lower axes)
    and as a function of the irradiation fluence (left axes). The solid white
    lines separate the regions with high enough defect concentration to form a
    ferromagnetic phase (FM) from those with low defect concentration forming a
    paramagnetic phase (PM) and indicates the percolation transition. The dashed
    white lines indicate the shift of the sample surface due to sputtering. At
    fluences $> 4\times 10^{15}$~cm$^{-2}$, the thickness of the FM phase is
    stable, as indicated by the arrows in (a,c,e). The visible steps in the
    color maps are due to the discretization of the depth, which has a step size
    of $0.25c = 2.4$~\AA, corresponding to the anatase layer spacing in the
    (001) crystal direction. }
\end{figure}

Using SRIM simulations for Ar$^+$ ions at $E_\mathrm{ion} = 200$~eV, $500$~eV
and $1000$~eV, and taking into account the defect creation probabilities
(Figure~\ref{fig:8}), a magnetic phase diagram can be constructed. At low
energies, the sample volume affected by the incident ions is much smaller than
in the high energy case and the effect of surface sputtering has to be taken
into account. SRIM allows to calculate the sputtering rate. We find values of
1.1, 1.5 and 2.3 nm / ($10^{16}$ ions / cm$^{2}$) at ion energies of 200~eV,
500~eV and 1000~eV, respectively.

Figure~\ref{fig:9} shows the dFP density and degree of amorphization as a
function of the Ar ion fluence for the three ion energies considered in this
study. Due to sputtering, the amorphization of the host lattice has a much
smaller effect, as the thin amorphous surface layer is constantly removed. This
is indicated by a dashed line in Figure~\ref{fig:9}. The solid line indicates
the magnetic percolation transition between unordered (paramagnetic) and FM
phases.

At fluences $> 4\times 10^{15}$~cm$^{-2}$, the defect creation and sputtering
processes are at equilibrium. The volume and defect densities of the emerging FM
phase stay constant over the whole fluence range. The equilibrium volume depends
on the ion energy, as indicated by the thickness of the FM regions along the
irradiation direction in Figure~\ref{fig:9}. At $E_\mathrm{ion} = 200$~eV, the
emerging FM layer grows to an equilibrium thickness $d_\mathrm{FM} = 4.6$~\AA,
corresponding to $0.48 c$ ($c=9.51$~\AA, the Anatase lattice constant). At
$E_\mathrm{ion} = 500$~eV we find an equilibrium thickness $d_\mathrm{FM} =
12.0$~\AA~($=1.26 c$) and at $E_\mathrm{ion} = 1000$~eV, $d_\mathrm{FM} =
19.9$~\AA~($=2.09 c$). The anatase unit cell consists of four layers, stacked
along the (001) crystal direction. Therefore, the emerging FM phase is expected
to be restricted to the first 2, 4 and 8 layers of the host lattice,
respectively.

\subsection{Experimental observation of the emerging FM phase in Anatase TiO$_2$}

\begin{figure}
  \centering
  \includegraphics[width=14.5cm]{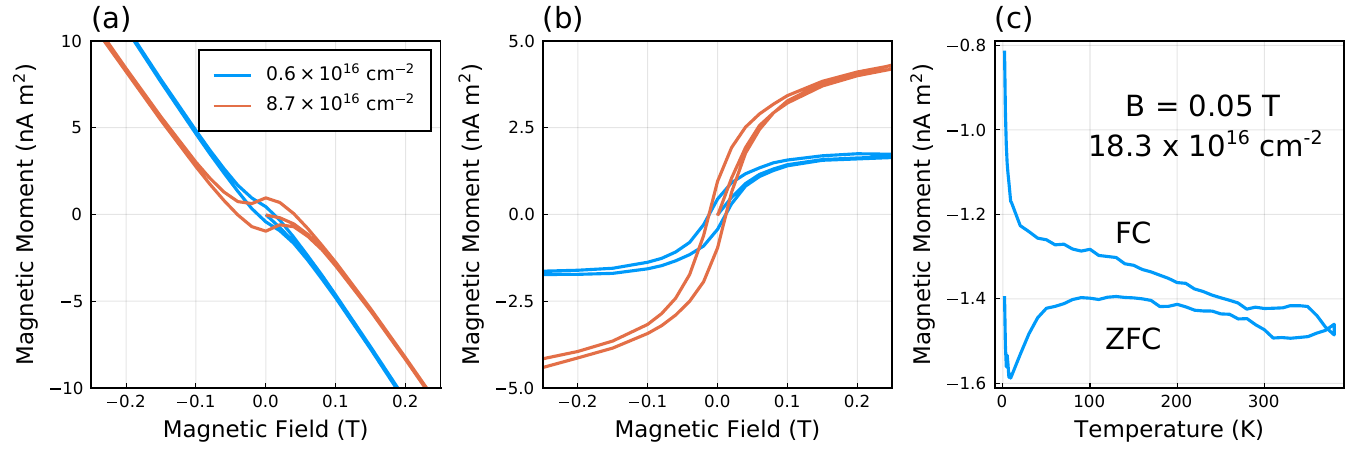}
  \caption{\label{fig:10} (a,b) TiO$_2$ thin film irradiated with $0.6 \times
    10^{16}$~cm$^{-2}$ (blue) and $8.7 \times 10^{16}$~cm$^{-2}$ (orange) Ar$^+$
    ions at $E_\mathrm{ion} = 1000$~eV. (a)~Hysteresis loop showing total
    magnetic moment as a function of applied magnetic field at $T = 300$~K.
    (b)~Hysteresis loop after subtracting a linear diamagnetic background. The
    magnetic field was applied parallel to the film surface. (c) Zero-field
    cooled (ZFC) / field cooled (FC) curve, measured at $B=0.05$~T for sample
    S1000 at an irradiation fluence $18.3\times 10^{16}$~cm$^{-2}$. }
\end{figure}

Figure~\ref{fig:10}(a) shows two typical hysteresis curves, measured at room
temperature and after sample S1000 had been irradiated with a fluence of
$0.6\times 10^{16}$~cm$^{-2}$ (blue) and $8.7\times 10^{16}$~cm$^{-2}$ (orange).
After subtracting the linear diamagnetic background (Figure~\ref{fig:10}(b)), a
hysteretic FM signal clearly appears, showing a magnetic moment at saturation
$m_\mathrm{sat}$ that increases with ion fluence (hysteresis curves at all
irradiation fluences are shown in Figures S1-S3 in the supporting information).
Figure~\ref{fig:10}(c) shows the zero-field cooled (ZFC) and field cooled (FC)
curves, measured at an irradiation fluence of $18.3\times 10^{16}$~cm$^{-2}$,
in the temperature range 2-380~K and at an applied magnetic field of $B=0.05$~T.
The opening between the ZFC and FC curves is a clear sign of a FM phase.

\begin{figure}
  \centering
  \includegraphics[width=8.5cm]{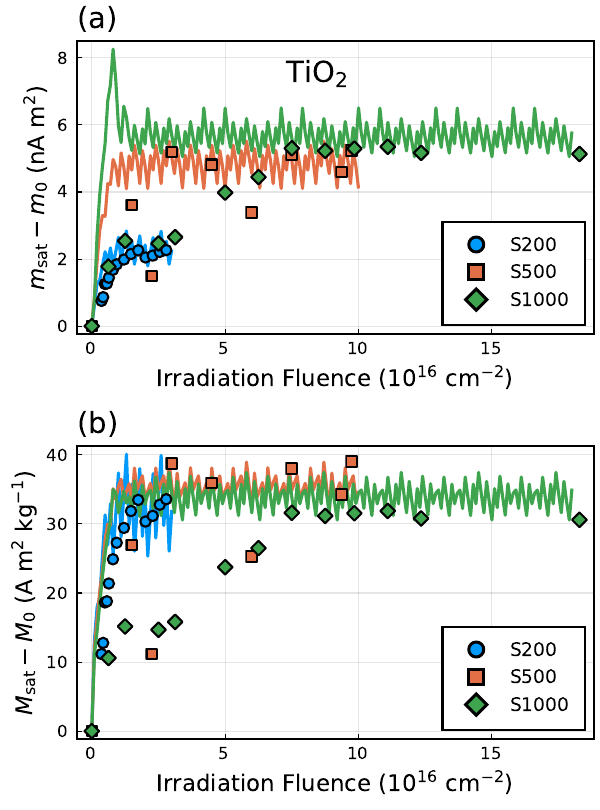}
  \caption{\label{fig:11} Magnetic moment $m_\mathrm{sat}$ (a) and magnetization
    $M_\mathrm{sat}$ (b) at saturation of the ferromagnetic signal, measured at
    room temperature as a function of the ion irradiation fluence, in the three
    TiO$_2$ thin film samples, S200 ($\bullet$), S500 ($\blacksquare$) and S1000
    ($\blacklozenge$). The background signal, $m_0$ ($M_0$) measured before any
    irradiation has been subtracted from the experimental data. The solid lines
    show the magnetic moment (magnetization) values predicted from our
    theoretical calculations (Figure~\ref{fig:9}). }
\end{figure}

By systematically measuring the magnetic hysteresis of the samples as a function
of the irradiation fluence, we can gain some insight into the evolution of the
emerging FM phase. Figure~\ref{fig:11}(a) shows the total magnetic moment at
saturation at $T=300$~K, $m_\mathrm{sat}$, after subtracting the linear
diamagnetic background, as a function of the irradiation fluence, for the three
samples S200, S500 and S1000. The values of $m_\mathrm{sat}$ have been obtained
by fitting the hysteresis curves with Equation~(\ref{eq:hysteresis_model}). The
background signal $m_0$, measured in each sample before any irradiation was
subtracted. The magnetization $M_\mathrm{sat}$ was calculated from the measured
total moment $m_\mathrm{sat}$ taking into account the equilibrium volume of the
magnetic phase predicted from the phase diagram (Figure~\ref{fig:9}).

We first observe that the moment at saturation increases with increasing total
irradiation fluence and saturates at high fluences. This is expected, as
increasing the irradiation fluence also increases the density of dFP defect
complexes until the defect creation and sputtering processes reach equilibrium,
in agreement with the phase diagram (Figure~\ref{fig:9}). The saturation
magnetization of all three samples reaches values of the order of
35~Am$^2$kg$^{-1}$, which corresponds to a mean dFP concentration of $\sim
17$~at.~\% or one dFP per unit cell.

By numerically integrating the dFP concentration found in our calculations over
the volume of the FM phase in Figure~\ref{fig:9} and taking into account the
magnetic moment ($2\mu_B$) of each dFP defect, we can calculate the expected
magnetization and total moment of the samples as a function of the irradiation
fluence. The result is shown in Figure~\ref{fig:11} as solid lines. The
oscillations are numerical artifacts due to the discretization of the phase
diagram. At the lowest ion energy (sample S200), our predictions show excellent
agreement with the experimental data. The predicted equilibrium magnetization at
high irradiation fluences of all three samples also agrees very well with our
measurements. The evolution of the magnetization of sample S1000 to the
equilibrium value, on the other hand, does not agree well with our calculations,
that predict a much steeper approach to equilibrium. In fact, it appears like
the saturation moment of sample S1000 first follows the same evolution as sample
S200 (Figure~\ref{fig:11}) and then slowly increases to the equilibrium value.

\subsection{The magnetic percolation process}

We have seen in Figure~\ref{fig:9} that the FM phase emerges in an ultrathin
region of thickness between 2 and 8 anatase layers. To better understand the
evolution of these ultrathin FM layers, it is instructive to take a closer look
at the magnetic percolation process, i.e. the transition from isolated local
magnetic moments to a long-ranged ordered phase upon increasing the defect
density.

A system of dilute defects in a host lattice that interact magnetically on a
finite length scale can be described using the framework of percolation
theory~\cite{Stauffer1994,Sahini1994}. In the site-percolation model, sites on
the host lattice can be occupied by a defect and one can associate a probability
$p$ with the occupation of a site. At $p=0$, no defects are present in the
system and at $p=1$, all sites are occupied by a defect. This probability is
naturally related to the defect density in the system through the density of
possible sites that can host a defect.

The length scale of the magnetic interaction, namely the exchange coupling,
determines whether two occupied sites are linked: Two occupied sites that are
close enough to each other to interact ferromagnetically through the exchange
interaction form a percolation domain. The magnetic moments associated to each
site belonging to one percolation domain are correlated, while those associated
to different percolation domains behave independently.

The site-percolation model describes a second order geometrical phase
transition, that occurs at a critical occupation probability $p_c$. At $p <
p_c$, in the subcritical regime, small isolated percolation domains form. The
microscopic magnetic moments within each of these percolation domains are
aligned parallel to each other due to the ferromagnetic coupling, but their
orientation is arbitrary. The summed magnetic moment of each domain can be
treated as one large paramagnetic center and the system is in a
super-paramagnetic phase. At $p > p_c$, in the supercritical regime, one large
percolation domain exists that spans over the whole system volume. This domain
is called the percolation continent and exhibits the features of a ferromagnet,
such as a spontaneous magnetization.

At criticality, when $p \sim p_c$, the percolation continent emerges and the
defective host system becomes ferromagnetic. As with all second order phase
transitions, the percolation transition can be described by an order parameter,
namely the probability $P_\infty$, that a random defect belongs to the
percolation continent. Near criticality, the evolution of the order parameter
follows a power law of the form
\begin{equation}
  \label{eq:power_law}
  P_\infty \propto (p-p_c)^\beta,
\end{equation}
with a universal critical exponent $\beta$, that only depends on the
dimensionality of the system.

\begin{figure}
  \centering
  \includegraphics[width=12.5cm]{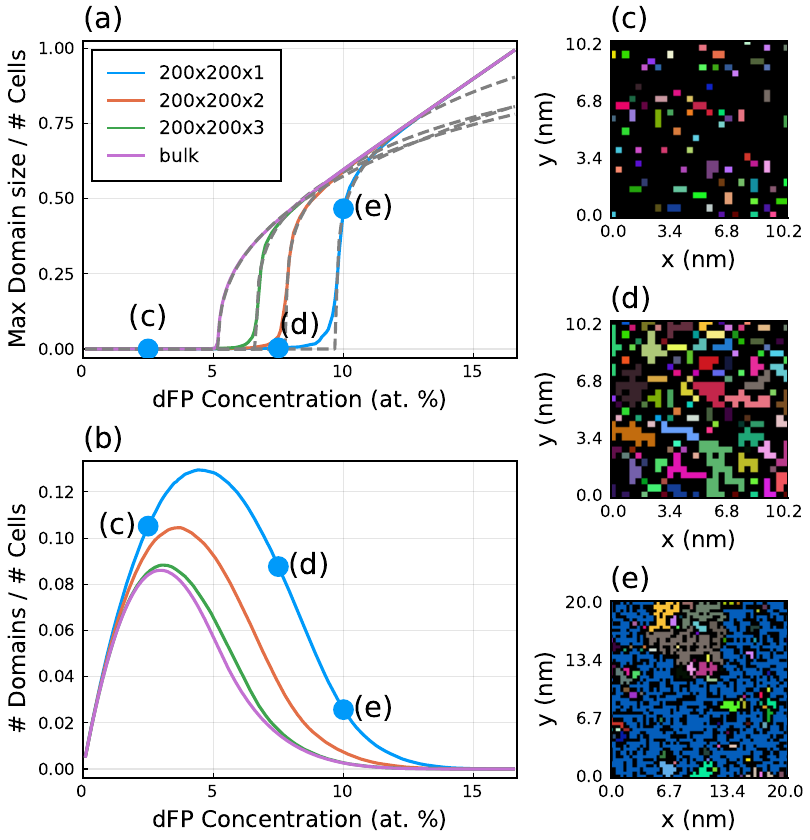}
  \caption{\label{fig:12} Magnetic percolation process in $200 \times 200 \times
    Z$ ($Z = 1, 2, 3$ and bulk) unit cell TiO$_2$ slabs for varying dFP
    concentrations. Nearest neighbor cells containing a dFP interact and form
    percolation domains. Varying the dFP concentration, three regimes can be
    seen: at low concentration ($< 5$~at.~\% in the monolayer), the sample is
    paramagnetic and the percolation domains are small. In the intermediate
    regime (5-9~at.~\% in the monolayer), the magnetic dFP defects start to
    interact and the domains grow rapidly in size. At the percolation threshold,
    9~at.~\%, 7.5~at.\%, 6~at.\% and 5~at.~\% in the mono-, bi-, trilayer and
    the bulk system, respectively, the domains merge and form a large
    percolation continent. The size of the percolation continent is shown in (a)
    as a function of the dFP concentration; (b) shows the number of percolation
    domains (normalized by the total number of cells). (c)-(e) Examples of
    monolayer grids, in which individual domains are color-coded. Black cells
    correspond to non-magnetic cells, that do not contain any dFP defect.
    Regions of the same color correspond to percolation domains, in which each
    cell has at least one nearest neighbor cell containing a dFP. The defect
    concentration was set to 2.5~at.~\% (c), 7.5~at.~\% (d) and 10~at.~\% (e).
    These three examples are marked by blue dots in (a) and (b). The dashed
    lines in (a) correspond to the fits of Equation~(\ref{eq:power_law}).}
\end{figure}

We simulate the magnetic percolation process in a grid of $200 \times 200$
anatase TiO$_2$ unit cells and a thickness of 1, 2 and 3 layers, enforcing
periodic conditions on the lateral boundaries. For comparison, we perform the
same simulations in a $200 \times 200 \times 200$ unit cell grid, enforcing
periodic boundary conditions in all three directions for a bulk system. Magnetic
dFP defects are randomly distributed throughout the grids, varying the total
defect concentration. Only nearest neighbor interactions between dFP defects are
taken into account, such that two nearest neighbor cells, each containing a dFP
defect, interact ferromagnetically and form a percolation domain.
Figure~\ref{fig:12}(a) shows the size of the largest of the domains, the
percolation continent, normalized to the total size of the grid.
Figure~\ref{fig:12}(b) shows the number of independent percolation domains
within the grid, as a function of the dFP defect concentration, normalized to
the total number of cells in the grids.

Varying the defect density, we see three regimes: at low concentration ($<
5$~at.~\% in the monolayer), the sample is paramagnetic and the percolation
domains are small (see Figure~\ref{fig:12}(c)). In the intermediate regime
(5-9~at.~\% in the monolayer), the magnetic dFP defects start to interact and
the domains grow rapidly in size (see Figure~\ref{fig:12}(d)). At the
percolation threshold, 9~at.~\%, 7.5~at.\%, 6~at.\% and 5~at.~\% in the mono-,
bi-, trilayer and the bulk system, respectively, the domains merge and the
percolation continent grows rapidly until spanning over the whole grid (see
Figure~\ref{fig:12}(e)). We note that at a dFP density of 17~at.~\%, which we
found at equilibrium in Figure~\ref{fig:11}, the percolation continent has fully
evolved and the order parameter $P_\infty = 1$.

\begin{table}[!t]
  \caption{Critical exponents $\beta$ obtained by fitting
    Equation~(\ref{eq:power_law}) to the data shown in Figure~\ref{fig:12}(a).}
  \label{tab:crit_exp}
  \centering
  \begin{tabular}{c|c}
    \# layers & $\beta$\\
    \hline
    1 & $0.210 \pm 0.007$\\
    2 & $0.220 \pm 0.006$\\
    3 & $0.259 \pm 0.007$\\
    bulk & $0.417 \pm 0.004$
  \end{tabular}
\end{table}

Table~\ref{tab:crit_exp} shows the critical exponents obtained by fitting the
curves shown in Figure~\ref{fig:12}(a) to Equation~(\ref{eq:power_law}).
Experimentally, the magnetic percolation transition can be obtained by taking
the remanent magnetic moment at zero applied magnetic field and at high
temperatures. Indeed, at ion fluences below the percolation transition, the
samples are expected to be (super-)paramagnetic and no remanence is expected
above the blocking temperature. Near the percolation transition, when the
percolation continent forms at a critical fluence $f_c$, the remanent magnetic
moment should follow the same critical behavior as the order parameter
$P_\infty$.

\begin{figure}
  \centering
  \includegraphics[width=12.5cm]{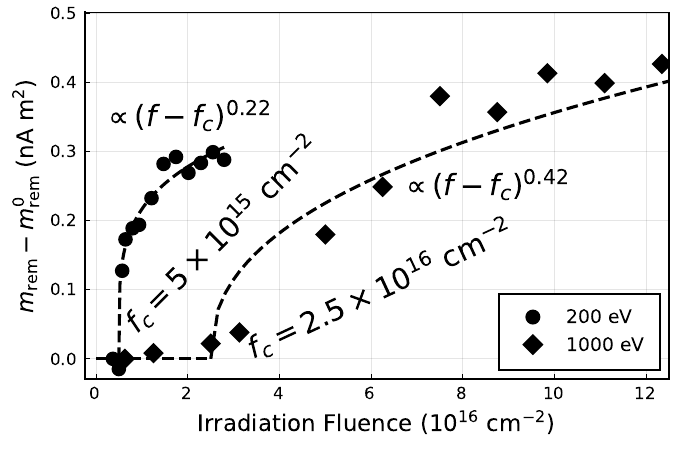}
  \caption{\label{fig:13} Remanent magnetic moment $m_\mathrm{rem}$ at zero
    field, measured at $T=300$~K after setting a magnetic field $B = 5$~T, as a
    function of the irradiation fluence $f$. The background remanence
    $m_\mathrm{rem}^0$ of the unirradiated samples was subtracted. The symbols
    represent experimental data of samples S200 ($\bullet$) and S1000
    ($\blacklozenge$). The dashed lines represent fits to $m_\mathrm{rem}
    \propto (f-f_c)^\beta$, with the critical fluences ($f_c$) and critical
    exponents ($\beta$) as indicated. }
\end{figure}

Figure~\ref{fig:13} shows the evolution of the remanent magnetic moment
$m_\mathrm{rem}$ measured in samples S200 and S1000 as a function of the
irradiation fluence $f$. By fitting the data to a power law ($m_\mathrm{rem}\propto
(f-f_0)^\beta$), we find critical fluences $f_0 = 5\times 10^{15}$~cm$^{-2}$ and
$2.5\times 10^{16}$~cm$^{-2}$ for samples S200 and S1000, respectively. The
resulting critical exponents are $\beta = 0.22 \pm 0.04$ and $0.42 \pm 0.07$,
respectively. Comparing these exponents to the theoretical values
(Table~\ref{tab:crit_exp}), we see that the remanence observed in sample S1000
follows the critical behavior of a bulk 3D percolation transition, while sample
S200 follows the critical behavior of a magnetic bilayer system. These results
match very well with the aforementioned thickness of the FM phases (see
Figure~\ref{fig:9}), that predicted the emergence of a magnetic bilayer in
sample S200, while in sample S1000 the FM phase spans over 8 layers.

\section{Dimensionality and surface effects, emergence of a perpendicular
  magnetic anisotropy}

Two dimensional long ranged magnetic order has long been thought to be
impossible at finite temperatures, as stated by the Mermin-Wagner (MW)
theorem~\cite{Mermin1966}: In bulk 3D ferromagnets, the exchange interaction
asserts a long range magnetic order up to the Curie temperature, $T_C$, where
thermal fluctuations become strong enough to randomize the spin orientation. In
2D magnetic systems with isotropic exchange interaction, the dimensionality
effect leads to an abrupt jump in the magnon dispersion and therefore strong
spin excitations at any finite temperature, destroying the magnetic order. The
presence of a strong uniaxial local magnetic anisotropy opens a gap in the
magnon dispersion, counteracting the Mermin-Wagner theorem in 2D and restoring
long range order. This has been demonstrated experimentally in ultrathin
transition metal films~\cite{Wuttig2004,Vaz2008} and 2D magnetic van der Waals
materials~\cite{Barzola2007,Gong2019,Wang2020}. 2D magnetic structures are not
only interesting from a fundamental physics perspective, but also regarding
their possible applications in 2D spintronics, magnonics or
spin-orbitronics~\cite{Chappert2007,Gong2019,Sharma2020,Hirohata2020,Wang2020}.
In the following section, we shall show the role of the magnetic anisotropy and
of the surface to stabilize the artificial FM phase at room temperature in
TiO$_2$ hosts, even in two dimensions.

\subsection{Measuring the magnetic anisotropy of the emerging FM phase in
  TiO$_2$ hosts}

\begin{figure}
  \centering
  \includegraphics[width=12.5cm]{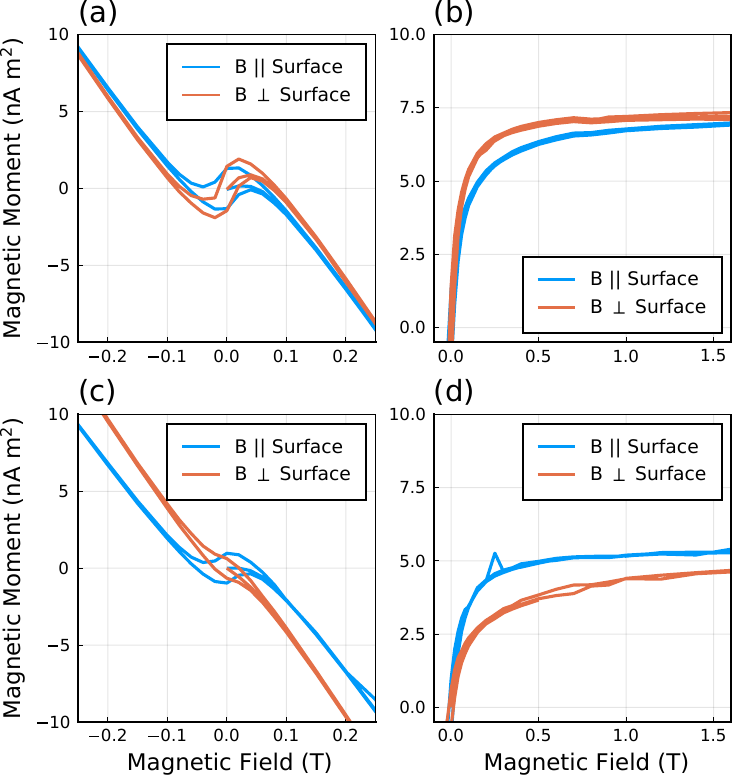}
  \caption{\label{fig:14} Magnetic hysteresis loops measured with the external
    magnetic field applied in the plane of the TiO$_2$ films (blue) and
    out-of-plane (orange), measured at $T=300$~K. The left panels show the total
    magnetic moment measured with the SQUID, the right panels show the
    ferromagnetic signal after subtracting the background diamagnetic and
    paramagnetic signals. Panels (a),(b) show the results obtained for sample
    S200 after irradiation with a fluence of $2.5\times 10^{16}$~cm$^{-2}$ and
    panels (c),(d) for sample S500 at a fluence of $6.0\times
    10^{16}$~cm$^{-2}$. }
\end{figure}

Figure~\ref{fig:14} shows measurements of the magnetic hysteresis loops obtained
by applying an external magnetic field parallel to the film surface (blue) and
perpendicular to the surface (orange) of the three TiO$_2$ samples. In panels
(a,c), the raw signals are shown and a clear magnetic anisotropy is visible.
After subtracting the linear diamagnetic contribution
(Figure~\ref{fig:14}(b,d)), the total magnetic anisotropy energy (MAE) can be
calculated as the area difference between the two curves. Here, we calculated
the area difference by first fitting Equation~(\ref{eq:hysteresis_model}) to the
experimental data and integrating the result analytically. To compensate
differences in the resulting saturation moments $m_\mathrm{sat}$ for the two
field orientations, e.g. due to fitting error or finite sample size effects (see
Section~IV.A), we rescaled the values of $m_\mathrm{sat}$ and $m_\mathrm{rem}$,
such that $m_\mathrm{sat}$ coincides. The sign is defined such that positive MAE
indicates a magnetic easy in-plane direction, parallel to the film surface while
a negative MAE indicates an out-of-plane easy axis. More details are given in
the supporting information. At the selected irradiation fluences, sample S200
(panels (a,b)) shows a perpendicular magnetic anisotropy while sample S500
(panels (c,d)) shows an in-plane anisotropy.

\begin{figure}[h!]
  \centering
  \includegraphics[width=12.5cm]{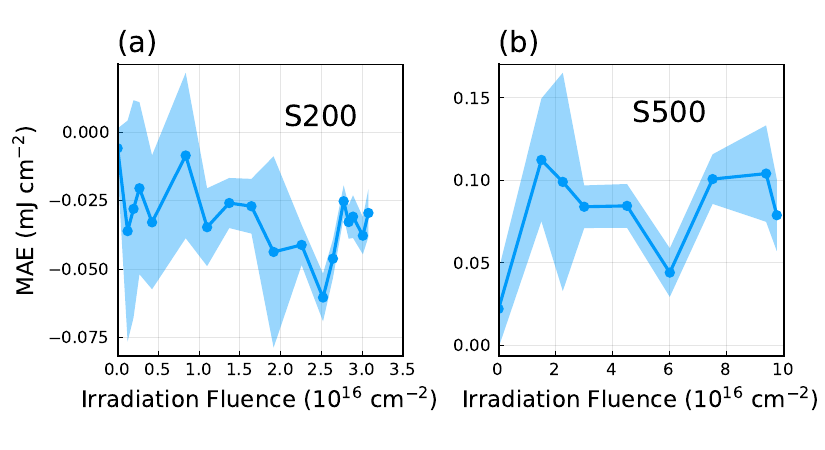}
  \caption{\label{fig:15} The magnetic anisotropy energy (MAE) obtained from the
    magnetic hysteresis curves measured after each irradiation step for the
    TiO$_2$ sample S200 (a) and S500 (b). The MAE is defined such that a
    negative value indicates an out of plane easy-axis (along the film surface
    normal), while a positive value indicates an easy in-plane magnetization
    direction (parallel to the film surface). The shaded area indicates the
    confidence margin within 5\% significance level. }
\end{figure}

Figure~\ref{fig:15} shows the total MAE as a function of the irradiation
fluence, of the samples S200 (a) and S500 (b). In sample S200, where the
thickness of the FM phase is estimated to only two layers of the host lattice,
the MAE is negative throughout all irradiation fluences, indicating a magnetic
easy axis normal to the film surface. In sample S500, the MAE is positive and
its magnitude is roughly four times larger than that of sample S200. For sample
S500, the magnetic phase diagram (Figure~\ref{fig:9}) predicts a FM phase
spanning the first four layers of the film surface, i.e. twice as many as for
sample S200, hinting towards the role of the surface in the emergence of a
perpendicular magnetic anisotropy.

\subsection{DFT electronic structure calculations of the defective
  T\MakeLowercase{i}O$_2$ surface}

\begin{figure}[h!]
  \centering
  \includegraphics[width=8.5cm]{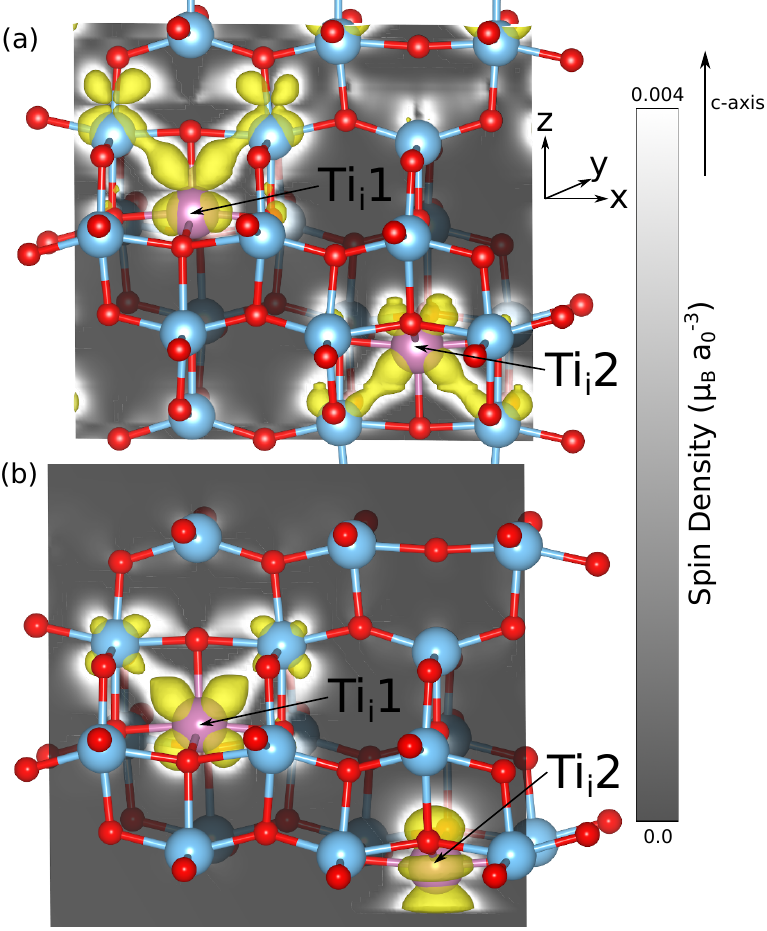}
  \caption{\label{fig:16} FLAPW-DFT calculations for (a) a 3x3x1 bulk anatase
    supercell and (b) a 3x3x1 supercell of the (001) anatase surface, each
    containing one dFP defect per supercell. The relaxed atom positions are
    represented by spheres (blue: Ti, red: O, purple: Ti$_i$). The isosurface at
    0.005 $\mu_B a_0^{-3}$ spin density is shown in yellow. The spin density in
    the (010)-plane through the two interstitials is indicated by shades of gray
    according to the scale on the right. }
\end{figure}

To understand the origin of the magnetic anisotropy shown in
Figure~\ref{fig:15}, we performed DFT electronic structure calculations using
the full potential linearized augmented plane wave (FLAPW) method implemented in
the FLEUR code, including spin-orbit interaction and a Hubbard term $U =
4.0$~eV, on a $3 \times 3 \times 1$ supercell of anatase TiO$_2$, containing one
dFP defect (the defect labeled ``di-FP1'' in Ref.~\cite{Stiller2020}). We used a
planewave cutoff times muffin tin radius $K_\mathrm{max} \times a_\mathrm{MT} =
7.0$. The atomic structure was relaxed using a $2 \times 2 \times 2$ k-point
grid and the final charge and spin density was calculated using a $6 \times 6
\times 6$ k-point grid. Figure \ref{fig:16}(a) shows the relaxed bulk atomic
structure, with the two Ti interstitials colored in pink. The isosurface at a
spin density of $0.005 \mu_B a_0^{-3}$ is shown in yellow and is mainly located
around the two interstitials and their neighboring Ti lattice atoms in the (010)
plane, having $d_{xz}$ character, as confirmed by the density of states
(Figure~\ref{fig:16}(a)). We find a total magnetic moment of $2\mu_B$/dFP
defect. These calculations match well with those presented by Stiller et
al.~\cite{Stiller2020}. We also calculated the MAE from the total energy
difference between the in-plane and out-of-plane magnetization state and find
$\mathrm{MAE} = 11$~$\mu$eV/atom with an (001) easy-plane in the bulk, as
opposed to the out-of-plane easy axis found by Stiller et al~\cite{Stiller2020}.

\begin{figure}[h!]
  \centering
  \includegraphics[width=8.5cm]{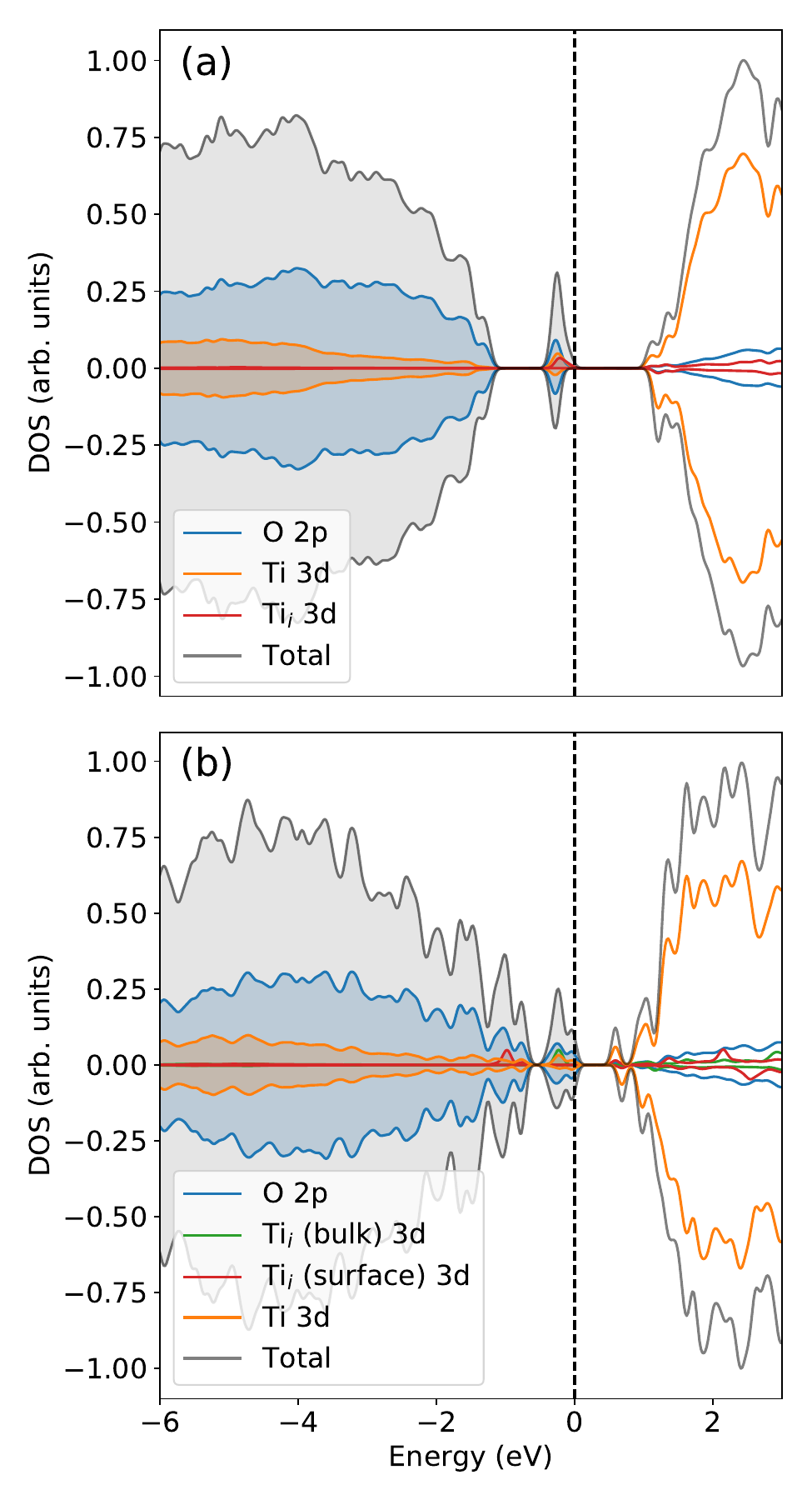}
  \caption{\label{fig:17} Density of states (DOS) (a) of the bulk anatase
    structure and (b) of the (001)-surface, each containing $5.5$\% dFP. The
    total DOS is shown in gray; the partial DOS (PDOS) of O 2p states is shown
    in blue; the PDOS of Ti 3d states is shown in orange. In (a), the PDOS of
    the two equivalent interstitial 3d states is shown in red; in (b), the PDOS
    of the bulk (Ti$_i$1) and surface (Ti$_i$2) interstitials are shown in green
    and red, respectively. }
\end{figure}

We then calculated the electronic structure of the dFP defect at the (001)
anatase surface, using a $3 \times 3$ supercell containing four layers of
anatase, as shown in Figure~\ref{fig:16}(b). Only the lower two layers were
relaxed, while the upper two were held fixed at the bulk atomic positions. After
structural relaxation, the surface layer shows displacements comparable to
values found in the literature~\cite{Araujo-Lopez2016} ($\alpha = 142^\circ$,
Ti$_{5c}$--O$_{2c}$ (short) $= 1.813$~\AA, Ti$_{5c}$--O$_{2c}$ (long) $=
2.010$~\AA, Ti$_{5c}$--O$_{3c} = 1.941$~\AA). As visible in
Figure~\ref{fig:16}(b), the spin density (shown in yellow) has similar structure
as in the bulk around the interstitial Ti$_i$1 on the second layer. On the first
layer, on the other hand, the spin density at interstitial Ti$_i$2 changes
strongly owing to the reduced coordination. There, the Ti 3d$_{z^2}$ orbital is
mainly spin polarized, as reflected by the DOS (Figure~\ref{fig:17}(b)). The
magnetic anisotropy at the (001) surface results in an out-of-plane easy axis
with a large $\mathrm{MAE} = -137$~$\mu$eV/atom.

These results together with the calculated defect distribution
(Figure~\ref{fig:9}) explain the measured MAE (Figure~\ref{fig:15}): At low ion
energy of $E_\mathrm{ion} = 200$~eV (sample S200), the FM phase emerges in the
first two layers at the film surface, resulting in a large negative MAE. At
higher ion energies, the FM phase emerges in a larger volume and also has
contributions from bulk states, which favor an in-plane easy-axis. Depending on
the local distribution of the defects, either the surface or the bulk states
dominate the total MAE.

\section{Conclusions}

The computational methods proposed in this work serve as a viable route toward
the systematic discovery of new artificial functional magnetic materials that
can be created experimentally by ion irradiation techniques. With a minimal
amount of input parameters, the scheme provides excellent quantitative
predictions in a large range of ion energies. The information gained from first
principles helps to understand existing experimental results and notably solve
the inherent problem of the experimental uncertainty regarding the magnetic
volume, which has been a major source of controversy.

By revisiting experimental results from the literature of a FM phase emerging in
SiC upon high energy ion irradiation and comparing them to our computational
predictions, we found that the main process limiting the evolution of the
artificial FM phase at high ion energies is the amorphization of the host
lattice.

In the case of low ion energies, sputtering of lattice atoms at the surface
plays an important role and limits the degree of amorphization, as we
demonstrated experimentally in anatase TiO$_2$ hosts. When the defect production
and sputtering processes reach an equilibrium, high defect densities (up to
17~at.~\%) can be created, which allows full magnetic percolation. We have shown
that at low enough ion energies ($E_\mathrm{ion} = 200$~eV in the TiO$_2$),
ultrathin ferromagnetic films down to a magnetic bilayer can be created.

In the ultrathin artificial FM layers created in TiO$_2$ hosts, we have
investigated the magnetic anisotropy and showed that a perpendicular magnetic
anisotropy emerges, depending on the thickness of the magnetic phase. We could
identify the origin of this PMA in the contribution of magnetic surface states,
as shown by DFT calculations.

\begin{acknowledgments}
  The authors thank A. Setzer and M. Stiller for fruitful discussions; Hichem
  Ben Hamed and Wolfram Hergert for the cooperation and support. Part of this
  study has been supported by the DFG, Project Nr. 31047526, SFB 762
  ``Functionality of oxide interfaces'', project B1. Computations for this work
  were done (in part) using resources of the Leipzig University Computing
  Centre.
\end{acknowledgments}

\appendix

\section{Magnetic Hysteresis Curves}

\begin{figure}[h!]
  \centering
  \includegraphics[width=12.5cm]{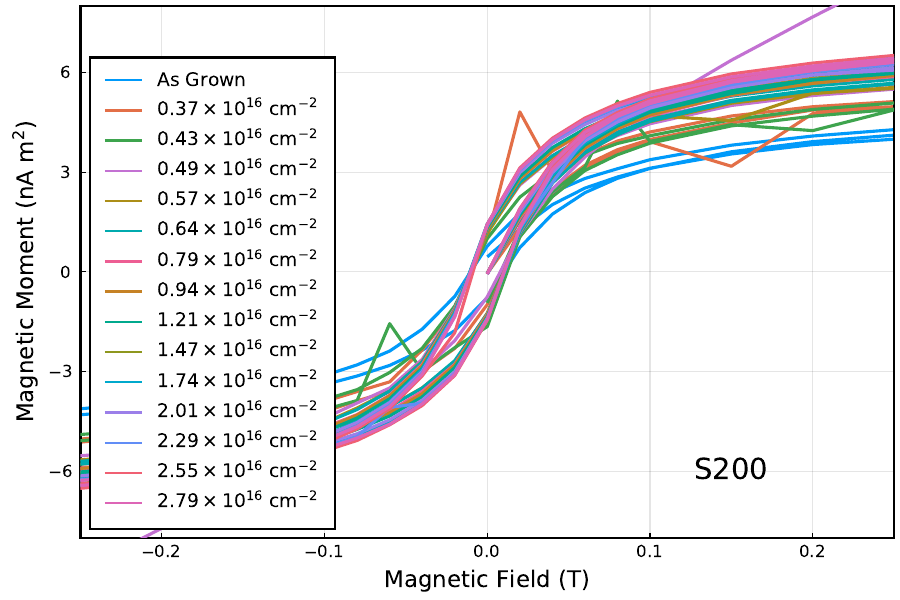}
  \caption{\label{fig:S1} Hysteresis loops measured in the TiO$_2$ sample S200
    at $T=300$~K after each Ar$^+$ irradiation step. The linear diamagnetic
    background was subtracted. }
\end{figure}

\begin{figure}[h!]
  \centering
  \includegraphics[width=12.5cm]{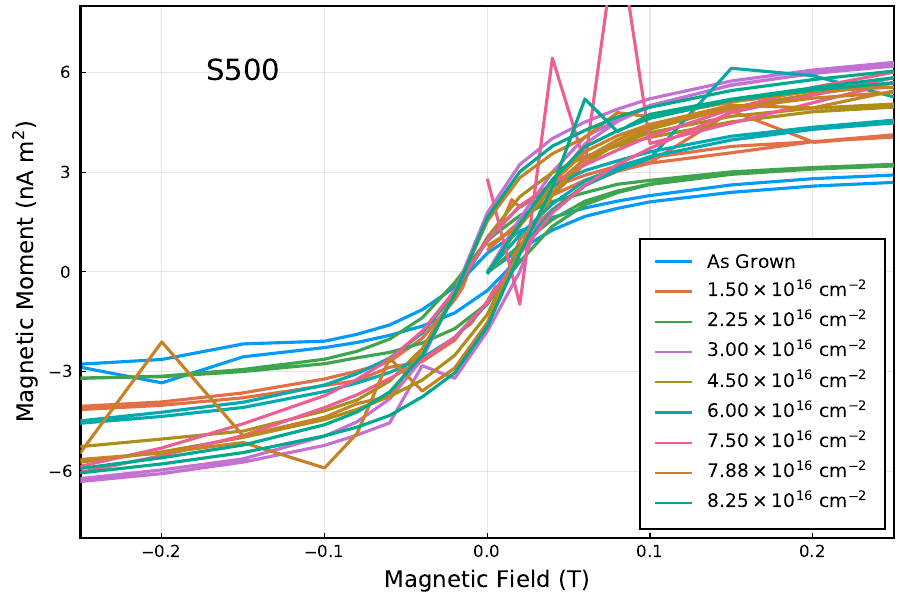}
  \caption{\label{fig:S2} Hysteresis loops measured in the TiO$_2$ sample S500
    at $T=300$~K after each Ar$^+$ irradiation step. The linear diamagnetic
    background was subtracted. }
\end{figure}

\begin{figure}[h!]
  \centering
  \includegraphics[width=12.5cm]{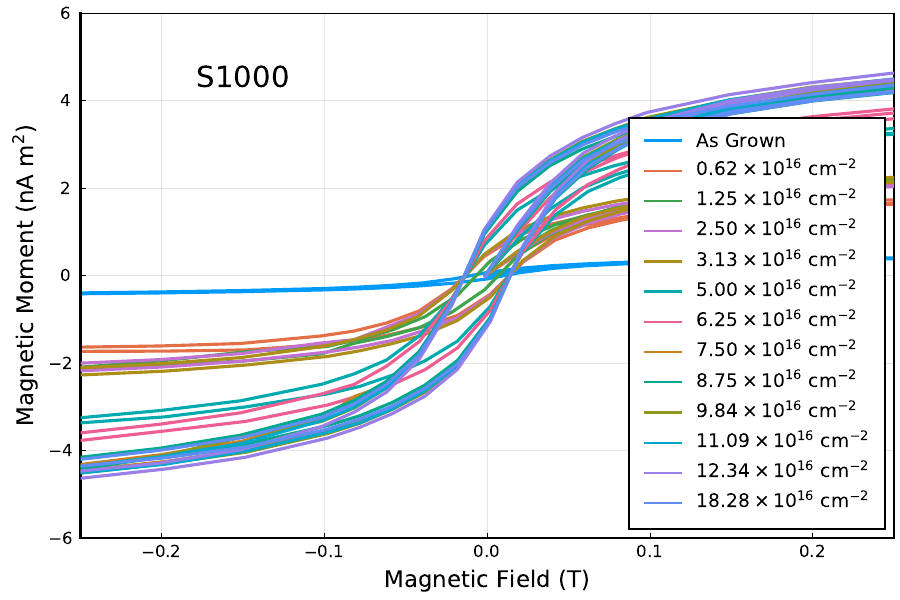}
  \caption{\label{fig:S3} Hysteresis loops measured in the TiO$_2$ sample S1000
    at $T=300$~K after each Ar$^+$ irradiation step. The linear diamagnetic
    background was subtracted. }
\end{figure}

\newpage
\section{Magnetic Anisotropy Energy}

\begin{figure}[h!]
  \centering
  \includegraphics[width=\textwidth]{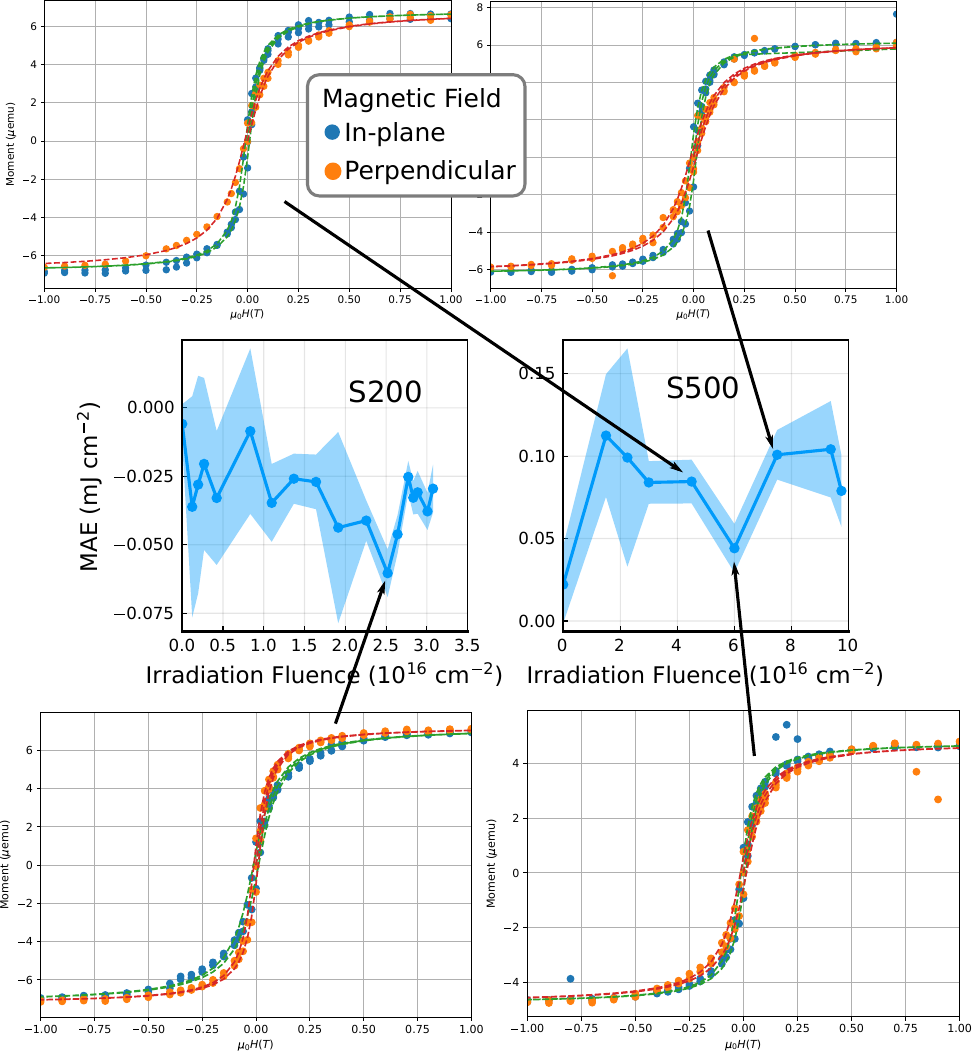}
  \caption{\label{fig:S4} Hysteresis loops, measured at $T=300$~K with a
    magnetic field applied in-plane and perpendicular to the film surface,
    corresponding to the magnetic anisotropy energy (MAE) values marked by
    arrows. The dashed lines show the fits to Equation~(1) in the main text. The
    method to obtain the MAE from the hysteresis curves is described in
    Section~V.A of the main text. }
\end{figure}

Figure~\ref{fig:S4} shows the magnetic anisotropy energy (MAE) of sample S200
and S500 as a function of the irradiation fluence. The MAE was obtained from
magnetic hysteresis curves such as those shown for some example fluences.
Hysteresis curves were measured once with the magnetic field applied in-plane
(blue bullets) and perpendicular to the sample surface (orange bullets). Each
curve was then fit using Equation~(\ref{eq:hysteresis_model2})
\begin{equation}
  \label{eq:hysteresis_model2}
  m(B; m_s, m_r, B_c) = \frac{2m_s}{\pi} \arctan\left[ \frac{B \pm B_c}{B_c} \tan\left( \frac{\pi m_r}{2m_s} \right) \right],
\end{equation}
to recover the saturation moment $m_s$, the remanent moment $m_r$ and the
coercive field $B_c$. The magnetic energy was then calculated along the virgin
curve as
\begin{equation}
  E(m_s, m_r, B_c) = \int_0^{B_\mathrm{max}} m(B; m_s, m_r, B_c) \mathrm{d}B.
\end{equation}
The integration cutoff $B_\mathrm{max}$ was set to 5~T. The MAE was then
calculated as
\begin{equation}
  MAE = \frac{1}{A}\left[ E(m_s^\parallel, m_r^\parallel, B_c^\parallel) - E(m_s^\parallel, m_r^\perp  m_s^\parallel / m_s^\perp, B_c^\perp) \right],
\end{equation}
with $A$ the sample surface area. We note that the saturation and remanent
moments obtained in the perpendicular configuration were rescaled, so that the
two saturation moments match.

\end{document}